\documentclass[sigconf,screen]{acmart}
\setcopyright{rightsretained}
\acmPrice{}
\acmDOI{10.1145/3611643.3613099}
\acmYear{2023}
\copyrightyear{2023}
\acmSubmissionID{fse23demo-p57-p}
\acmISBN{979-8-4007-0327-0/23/12}
\acmConference[ESEC/FSE '23]{Proceedings of the 31st ACM Joint European Software Engineering Conference and Symposium on the Foundations of Software Engineering}{December 3--9, 2023}{San Francisco, CA, USA}
\acmBooktitle{Proceedings of the 31st ACM Joint European Software Engineering Conference and Symposium on the Foundations of Software Engineering (ESEC/FSE '23), December 3--9, 2023, San Francisco, CA, USA}
\received{2023-05-11}
\received[accepted]{2023-07-20}

\usepackage[T1]{fontenc}
\usepackage{amsmath,amsfonts}

\usepackage{enumitem}

\usepackage{xspace}
\usepackage{listings}
\usepackage{url}

\usepackage[capitalise]{cleveref}
\usepackage{breakurl}
\usepackage{microtype}
\usepackage[shortcuts]{extdash}

\definecolor{dkgreen}{rgb}{0,0.6,0}
\definecolor{gray}{rgb}{0.5,0.5,0.5}
\definecolor{mauve}{rgb}{0.58,0,0.82}
\definecolor{shadecolor}{rgb}{0.95,0.95,0.95}

\definecolor{pblue}{rgb}{0.13,0.13,1}
\definecolor{pgreen}{rgb}{0,0.5,0}
\definecolor{pred}{rgb}{0.9,0,0}
\definecolor{pgrey}{rgb}{0.46,0.45,0.48}
\PassOptionsToPackage{usenames,dvipsnames}{color}

\newcounter{dgcounter}
\newcommand{\newdg}[2]{\noindent\refstepcounter{dgcounter}{\em #2} (\textbf{DG\arabic{dgcounter}):}\label{#1}}
\newcommand{\dgref}[1]{\textbf{DG\ref{#1}}}

\lstdefinestyle{toplisting}{
  float=tp,
  floatplacement=tbp,
}

\AtBeginDocument{%
  \providecommand\BibTeX{{%
    \normalfont B\kern-0.5em{\scshape i\kern-0.25em b}\kern-0.8em\TeX}}}

\begin{document}

\title[MASC: A Tool for Evaluating Crypto-API Misuse Detectors]{MASC: A Tool for Mutation-Based Evaluation of \\Static Crypto-API Misuse Detectors}

\author{Amit Seal Ami}\email{aami@wm.edu}
\orcid{0000-0002-9455-2230}
\authornote{These authors contributed equally to this paper}
\affiliation{
  \institution{Computer Science Department, William \& Mary}
  \streetaddress{}
  \city{Williamsburg}
  \state{Virginia}
  \country{USA}
  \postcode{}
}

\author{Syed Yusuf Ahmed}
\orcid{0009-0002-3229-1386}
\email{bsse1013@iit.du.ac.bd}
\authornotemark[1]
\author{Radowan Mahmud Redoy}
\orcid{0009-0002-8014-8023}
\email{bsse1002@iit.du.ac.bd}
\authornotemark[1]
\affiliation{%
  \institution{Institute for Information Technology, University of Dhaka}
  \streetaddress{}
  \city{Dhaka}
  \state{}
  \country{Bangladesh}
}

\author{Nathan Cooper}
\orcid{0000-0003-2498-705X}
\email{nacooper01@wm.edu}
\author{Kaushal Kafle}
\orcid{0000-0003-1917-7677}
\email{kkafle@wm.edu}
\affiliation{
  \institution{Computer Science Department, William \& Mary}
  \streetaddress{}
  \city{Williamsburg}
  \state{Virginia}
  \country{USA}
  \postcode{}
}

\author{Kevin Moran}
\email{kpmoran@ucf.edu}
\orcid{0000-0001-9683-5616}
\affiliation{%
  \institution{Department of Computer Science, University of Central Florida}
  \city{Orlando}
  \state{Florida}
  \country{USA}}

\author{Denys Poshyvanyk}
\email{denys@cs.wm.edu}
\orcid{0000-0002-5626-7586}
\affiliation{
  \institution{Computer Science Department, William \& Mary}
  \streetaddress{}
  \city{Williamsburg}
  \state{Virginia}
  \country{USA}
  \postcode{}
}

\author{Adwait Nadkarni}
\orcid{0000-0001-6866-4565}
\email{apnadkarni@wm.edu}
\affiliation{
  \institution{Computer Science Department, William \& Mary}
  \streetaddress{}
  \city{Williamsburg}
  \state{Virginia}
  \country{USA}
  \postcode{}
}

\renewcommand{\shortauthors}{Amit et al.}

\begin{abstract}
While software engineers are optimistically adopting crypto-API misuse
detectors (or crypto-detectors) in their software development
cycles, this momentum must be accompanied by a rigorous
understanding of crypto-detectors' \textit{effectiveness at finding crypto-API misuses in practice}.
This demo paper presents the technical details and usage scenarios of our tool, namely \textbf{M}utation \textbf{A}nalysis for evaluating \textbf{S}tatic \textbf{C}rypto-API misuse detectors (MASC).
We developed $12$ generalizable, usage based mutation operators and three mutation scopes, namely \textit{Main Scope}, \textit{Similarity Scope}, and \textit{Exhaustive Scope}, which can be used to expressively instantiate compilable variants of the crypto-API misuse cases.
Using MASC, we evaluated nine major crypto-detectors, and discovered $19$ unique, undocumented flaws.
We designed MASC to be \textit{configurable} and \textit{user-friendly}; a user can configure the parameters to change the nature of generated mutations. Furthermore, MASC comes with both Command Line Interface and Web-based front-end, making it practical for users of different levels of expertise.\\
Code: \url{https://github.com/Secure-Platforms-Lab-W-M/MASC}\\
\end{abstract}



\maketitle

\newcommand\codel[1]{\begin{verbatim}{#1}\end{verbatim}}

\newcommand\inline[1]{{\lstinline[keywordstyle=\color{black},basicstyle=\scriptsize\ttfamily,stringstyle=\color{black}]{#1}}}
\newcommand\inlinesmall[1]{{\lstinline[keywordstyle=\color{black},basicstyle=\small\ttfamily,stringstyle=\color{black}]{#1}}}

\newcommand\myparagraph[1]{\noindent\underline{\bf {#1}:}}
\newcommand\myparagraphnew[1]{\noindent{\bf {#1}:}}

\newcommand{\arrow}{{$\rightarrow$}\xspace}
\newcommand{\ie}{\textit{i.e.,}\xspace}
\newcommand{\eg}{\textit{e.g.,}\xspace}
\newcommand{\etc}{\textit{etc.}\xspace}
\newcommand{\etal}{\textit{et al.}\xspace}
\newcommand{\etals}{\textit{et al.'s}\xspace}
\newcommand{\aka}{{\small{a.k.a.}\xspace}}
\newcommand{\mascengine}{{\small{\tool{} Engine}\xspace}}
\newcommand{\masclab}{{\small{\tool{} Lab}\xspace}}
\newcommand{\pluginmanager}{{\small{Plugin Manager}\xspace}}
\newcommand{\configurationmanager}{{\small {Configuration Manager}\xspace}}
\newcommand{\TODO}[1]{{\color{red}{\textbf{TODO: {#1}}}}}
\newcommand{\REFH}[0]{{\color{red}REF}}
\newcommand{\bnum}[1]{{($#1$)}}
\makeatletter
\newcommand{\linebreakauthor}{%
  \end{@IEEEauthorhalign}
  \hfill\mbox{}\par
  \mbox{}\hfill\begin{@IEEEauthorhalign}
}
\newcommand{\tool}{{\small MASC}\xspace}
\newcommand{\tools}{{\small MASC}'s\xspace}
\newcommand{\mascweb}{{\small{\tool{} Web}\xspace}}
\newcommand{\mse}{$\mu$SE\xspace}
\newcommand{\mses}{$\mu$SE's\xspace}
\newcommand{\mdroid}{{\small MDroid$+$\xspace}}
\newcommand{\crashscope}{{\small CrashScope}\xspace}
\newcommand{\detector}{crypto\-/detector\xspace}
\newcommand{\detectors}{crypto\-/detectors\xspace}
\newcommand{\toolnamefull}{\textbf{M}utation \textbf{A}nalysis for evaluating \textbf{S}tatic \textbf{C}rypto-API misuse detectors (MASC)\xspace}
\newcommand{\cipher}{{Cipher}\xspace}
\newcommand{\ciphergetinstance}{\texttt{\small Cipher.getInstance(<parameter>)}\xspace}
\newcommand{\messageDigestInstance}{\texttt{\small MessageDigest.getInstance(<parameter>)}\xspace}
\newcommand{\messageDigest}{\texttt{\small MessageDigest}\xspace}
\newcommand{\des}{\texttt{des}\xspace}
\newcommand{\DES}{\texttt{DES}\xspace}
\newcommand{\AES}{\texttt{AES}\xspace}
\newcommand{\ECB}{\texttt{ECB}\xspace}
\newcommand{\MDFIVE}{\texttt{MD5}\xspace}
\newcommand{\muse}{$\mu$SE\xspace}

\newcommand{\trustManager}{\texttt{\small TrustManager}\xspace}
\newcommand{\xtrustManager}{\texttt{\small X509TrustManager}\xspace}
\newcommand{\extendedxtrustManager}{\texttt{\small X509ExtendedTrustManager}\xspace}
\newcommand{\getAcceptedIssuers}{\texttt{\small getAcceptedIssuers}\xspace}
\newcommand{\checkServerTrusted}{\texttt{\small checkServerTrusted}\xspace}
\newcommand{\checkClientTrusted}{\texttt{\small checkClientTrusted}\xspace}

\newcommand{\certificateException}{\texttt{\small CertificateException}\xspace}

\newcommand{\hostnameVerifier}{{\small HostnameVerifier}\xspace}
\newcommand{\trycatch}{\texttt{\small try-catch}\xspace}

\newcommand{\ivparameterspec}{\texttt{IvParameterSpec}\xspace}

\newcommand{\cryptoguard}{CryptoGuard\xspace}
\newcommand{\cryptoguardmultidex}{{$2,709$}\xspace}
\newcommand{\cryptoguarddownload}{{$4,353$}\xspace}
\newcommand{\cryptoguardtotal}{{$6,181$}\xspace}
\newcommand{\cryptoguardpercent}{{$62.23$\%}\xspace}
\newcommand{\cryptoguardandroiddottotal}{{$673$}\xspace}
\newcommand{\cryptoguardandroiddot}{{$383$}\xspace}
\newcommand{\cryptoguardendswithandroid}{{$290$}\xspace}
\newcommand{\cryptoguardandroiddottotalpercent}{{$10.89$\%}\xspace}
\newcommand{\androiddot}{{\texttt{android.}}\xspace}

\newcommand{\totalmisuses}{{$105$}\xspace}
\newcommand{\totaloperators}{{$12$}\xspace}

\newcommand{\crysl}{CrySL\xspace}
\newcommand{\cognicrypt}{CogniCrypt\xspace}
\newcommand{\xanitizer}{Xanitizer\xspace}
\newcommand{\coverity}{Tool$_X$\xspace}
\newcommand{\spotbug}{SpotBugs\xspace}
\newcommand{\spotbugfull}{SpotBugs with FindSecBugs\xspace}
\newcommand{\qark}{QARK\xspace}
\newcommand{\qpid}{Apache Qpid Broker-J\xspace}
\newcommand{\shiftleft}{ShiftLeft\xspace}
\newcommand{\codeqlgcs}{Github Code Security\xspace}
\newcommand{\codeqllgtm}{LGTM\xspace}

\newcommand{\coverityshort}{TX\xspace}
\newcommand{\xanitizershort}{XT\xspace}
\newcommand{\cryptoguardshort}{CG\xspace}
\newcommand{\shiftleftshort}{SL\xspace}
\newcommand{\qashort}{QA\xspace}
\newcommand{\codeqlgcsshort}{GCS\xspace}
\newcommand{\codeqllgtmshort}{LGTM\xspace}
\newcommand{\cryslshort}{CL\xspace}
\newcommand{\sportbugsshort}{SB\xspace}
\newcommand{\cognicryptshort}{CC\xspace}

\newcommand{\countflaws}{{$19$}\xspace}
\newcommand{\countFlawClasses}{{$5$}\xspace}
\newcommand{\countFlawClassesText}{{five}\xspace}

\newcommand{\no}{\remove{\ding{55}}}
\newcommand{\ye}{\ding{51}}
\newcommand{\pa}{\LEFTcircle}
\newcommand{\na}{\texttt{-}}
\newcommand{\nl}{\text{\O}}

\newcommand{\fcincomplete}{Flaw Class 1 (FC1): Incomplete Analysis of Target Code\xspace}
\newcommand{\fcdifferentcase}{Flaw Class 1 (FC1): String Case Mishandling\xspace}
\newcommand{\fcvalueresoluion}{Flaw Class 2 (FC2): Incorrect Value Resolution\xspace}
\newcommand{\fcomplexinheritance}{Flaw Class 3 (FC3): Incorrect Resolution of Complex Inheritance and Anonymous Objects\xspace}
\newcommand{\fcgenericnoise}{Flaw Class 4 (FC4): Insufficient Analysis of Generic Conditions in Extensible Crypto-APIs\xspace}%
\newcommand{\fcspecificnoise}{Flaw Class 5 (FC5): Insufficient Analysis of Context-specific, Conditions in Extensible Crypto-APIs\xspace}

\newcommand{\totalMutantApplications}{{$27$}\xspace}
\newcommand{\totalMutations}{{$20,303$}\xspace}
\newcommand{\totalMutationsAndroid}{{$2,515$}\xspace}
\newcommand{\totalMutationsJava}{{$17,788$}\xspace}
\newcommand{\totalReachMutant}{{$20,165$}\xspace}
\newcommand{\testedApps}{{$17$}\xspace}
\newcommand{\testedAndroidApps}{{$13$}\xspace}
\newcommand{\testedJavaComponents}{{$4$}\xspace}
\newcommand{\testedCryptoApps}{{$7$}\xspace}
\newcommand{\totalToolsUsed}{{$9$}\xspace}
\newcommand{\totalToolsUsedText}{{nine}\xspace}

\newcommand{\mainScope}{main\xspace}
\newcommand{\similarityScope}{similarity scope\xspace}
\newcommand{\exhaustiveScope}{exhaustive scope\xspace}
\newcommand{\totalMisuseCases}{{$105$}\xspace}
\newcommand{\newMisuseCases}{{$4$}\xspace}
\newcommand{\implementedMisuseCases}{{$19$}\xspace}

\newcommand{\dexlib}{{\texttt{dexlib2}}\xspace}

\section{Introduction}\label{sec:intro}
Software engineers have been relying on \detectors for decades to ensure the correct use of cryptographic APIs in the software and services they create, develop, and maintain~\cite{BBC+10}.
Such \detectors{} are ubiquitous in software engineering, as they are integrated into IDEs (\eg{} \cognicrypt plugin for Eclipse IDE~\cite{cognicrypteclipse}), testing suite of organizations such as Oracle Corporation~\cite{cryptoguard_oracle,rxa+19}, and for Continuous Integration/Continuous Deployment (CI/CD) pipelines~\cite{XanitizerRIGSIT,lgtm}.
In addition, hosting providers such as GitHub are formally provisioning such \detectors \eg GitHub Code Scan Initiative~\cite{github_third_party_code_scanners}.
In other words, the security of software and services are increasingly becoming more reliant on \detectors{}.
However, we have been relying on manually-curated benchmarks for evaluating the performance of \detectors, such benchmarks are known to be incomplete, incorrect, and impractical to maintain~\cite{owasp:mislabel}. Therefore, determining the effectiveness of \detectors{} from a \textit{security-focused perspective} requires a reliable and evolving evaluation technique that can scale with the volume and diversity of crypto-API and the different patterns of misuse.

We contextualized mutation testing techniques to create the \toolnamefull{} framework.
In our original, prototype implementation of \tool{}~\cite{ami-masc-oakland22}, it internally leveraged $12$ generalizable, usage-based mutation operators to instantiate mutations of crypto-API misuse cases for Java. The mutation operators were designed based on the design principles of Java Cryptographic Architecture (JCA)~\cite{JavaCryptographyArchitecture} and a threat model that consisted of users of varying skills and intentions (Section~\ref{sec:mutation-operators}).
\tool{} \textit{injects} these mutated misuse cases in Java or Android-based apps at three mutation scopes (injection sites), namely \textit{Similarity Scope} (extended from \mdroid~\cite{mtb+18,lbt+17}), \textit{Exhaustive Scope} (extended from \mse{}~\cite{demo-muse-2021,bkm+18,AKM+21}), and its independently developed \textit{Main Scope}, thus creating mutated applications that contain crypto-API misuse.
We demonstrated the practicality of prototype implementation of \tool{} by evaluating \totalToolsUsedText{} \detectors{} from industry and academia, and discovered \countflaws{} previously undocumented, unknown flaws that compromise the within-scope soundness of \detectors{}.
The full details of \tools{} methodology, design considerations, evaluation of \detectors leading to finding novel flaws, practical impact of found flaws in open source applications (therefore, the applicability of the mutation operators), and discussion of the findings are available in the original research paper~\cite{ami-masc-oakland22}.

In this paper, we present a mature implementation of \tool{} framework with focus on extensibility, ease of use, and maintainability to the stakeholders of \detectors, such as security researchers, developers, and users.
To elaborate, because of the newly developed plug-in architecture, \tool{} users can now create their own mutation operators that can be easily plugged into \tool{}, without diving deep into the existing code base ($11K+$ source lines of code).
Moreover, whereas the original prototype implementation of \tool{} involved semi-automated evaluation of \detectors{}, we made \tools{} workflow automated by leveraging the \textit{de-facto} SARIF~\cite{SARIF} formatted output of~\detectors{}.
Furthermore, we have created a web-based front-end of \tool{}'s implementation for the users to reduce the barrier to entry.
Finally, we restructured and refactored the open-source codebase of \tool{} to increase maintainability and extensibility of \tool{}, which will make future contributions and enhancements easier for both developers and open-source enthusiasts of \tool{}.
With these additions and enhancements, we hope that the current, open-source implementation of \tool{} will be used in finding flaws in, and thus helping to improve, existing \detectors.

\myparagraphnew{Contribution}
We present \tool{}, a user-friendly framework that leverages mutation-testing techniques for evaluating \detectors{}, with details of underlying techniques, design considerations, and improvements.
The new, key features of \tool{} are as follows:
\noindent\textit{Automated Evaluation of Crypto-detectors:} \tool{} can be used to evaluate \detectors{} in an end-to-end automated workflow within the Main Scope.

\noindent\textit{Customizable Evaluation of Crypto-detectors:} A user can customize the evaluation of \detectors{} by specifying the mutation operators for creating crypto-API misuse instances.

\noindent\textit{Plug-in Architecture for Custom Operators:}
\tool{} helps security researchers, developers and users, jump right into evaluating \detectors{} by creating their own, custom mutation operators that can be directly plugged-in the \textit{Main} Scope, without requiring them to learn and understand about the internal details of \tool{}.

\noindent\textit{User-friendly Front-end for End-users:} In addition to enhancing the command line interface of the original prototype implementation, we create and introduce an open-sourced, web-based front-end for end-users that can be run locally. The front-end contains an additional \textit{play-test-learn} interface, \masclab{}, where stakeholders can interact with mutation operators and can learn about mutating crypto-API misuse.

\myparagraphnew{Tool and Data Availability} The prototype implementation of the \tool{} framework, scripts and results of evaluating \detectors, as described in the original paper~\cite{ami-masc-oakland22}, are available in the \tool{} Artifact~\cite{MASCARTIFACT}.
Furthermore, the codebase of actively maintained, mature implementation of \tool{} is available separately with extensive documentation and examples~\cite{MASC}.
\section{Overview of MASC}\label{sec:overview}
\begin{figure}[tbp]
	\centering
    \includegraphics[width=.47\textwidth]{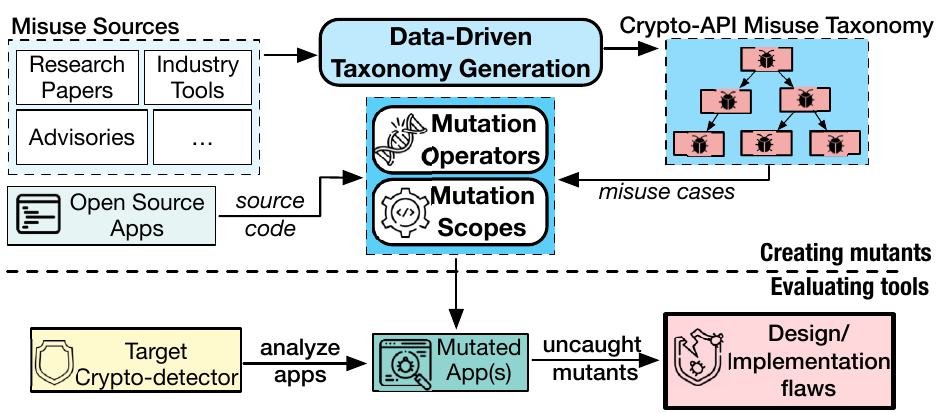}
    \caption{\small A conceptual overview of the \tool framework.}
    \label{fig:tool}
\end{figure}

Overall, \tool{} works by \bnum{1} mutating a base crypto API misuse case to create mutated crypto-API instantiations or mutated misuse case, \bnum{2} seeding or injecting the mutated misuse case in source code, \bnum{3} analyzing both unmutated and mutated source code using a target \detector{}, and \bnum{4} comparing the outputs of \detector applied on both base misuse case and mutated misuse case to identify undetected (not killed) mutated misuse case.
The overview of this process is shown in Figure~\ref{fig:tool}.

Conceptually, \tool{} contextualizes the traditional mutation testing techniques of SE domain for the evaluation of \detectors{}, while introducing \textit{crypto-API misuse mutation operators} that instantiates variants or expressions of crypto-API misuse.
\begin{lstlisting}[basicstyle=\ttfamily\scriptsize,float,numbers=none,caption={{\small
    Example crypto-API misuse instances created by \tool{}}},belowcaptionskip=-5mm,label=lst:mutations,emph={mutation},emphstyle=\bfseries,language=Java]
//base crypto API misuse
Cipher.getInstance("DES"); // 1
//mutated misuse instances from several mutation operators of MASC
Cipher.getInstance("des"); // 2
Cipher.getInstance("des".toUppercase()); //3
Cipher.getInstance("DE$S".replace("$","")); // 4
String val = "DES"; Cipher.getInstance(val); // 5
\end{lstlisting}

To elaborate, while mutation operators from the traditional, SE mutation testing are used to describe operations that either add, modify, or remove existing source code statement(s), in the context of \tool{}, \textit{crypto-API mutation operators} create expressive instances of crypto-API misuse independent of any source code or application.
As shown in Listing~\ref{lst:mutations}, statement marked //$1$ is the base misuse case, whereas statements //$2$ -- //$5$ are the mutated crypto-misuse cases instantiated by several mutation operators of \tool{}.
We provide the design considerations and implementation details of \tools{} mutation operators in Sec.~\ref{sec:mutation-operators}.
These mutated misuse instances are then "injected" or "seeded" in source code, where the injection site depends on the \textit{mutation scopes} of \tool{}, which we detail in Sec.~\ref{sec:mutation-scopes}.

\section{Design Goals}\label{sec:design-goals}
We considered several goals while designing \tool{}, while leaning on the experience we gained from the original version.

\newdg{1}{Diversity of Crypto-APIs}
Effectively evaluating \detectors{} requires considering misuse cases of existing crypto-APIs, which is challenging as crypto-APIs are as vast as the primitives they enable.
To address this, the crypto-API mutation operators need to be decoupled from the crypto-APIs.
Such implementation would mean that even in the case when new crypto APIs are introduced, \tool{} can still create mutated misuse cases as long as the new crypto-APIs follow existing design principles.

\newdg{2}{Open to Extension} While both original and current implementations of \tool{} come with $12$ generalizable mutation operators, these represent a subset of different expressions of misuse cases.
Hence, \tool{} should be open to extension by stakeholders so that they can create their own mutation operators that can be easily plugged-in to \tool{}, without needing to modify \tool{}.

\newdg{3}{Ease of Evaluating Crypto-detectors} While the original, semi-automated implementation of \tool{} required manual evaluating the target \detector{}, such heavy-lifting manual effort can not be simply expected from end-users.
Part of this manual effort was \textit{unavoidable} due to the unique, varied outputs produced by \detectors{}.
However, with the recent focus on using \detectors with CI/CD pipelines and the introduction of the \textit{de-facto} {\small SARIF}~\cite{SARIF} formatted outputs, it would become possible to not only automate the entire evaluation process, but also make it customizable.

\newdg{4}{Adapting to Users} Finally, \tool{} should be created in such a way that it is usable by users of varying skills and in different environments.
For instance, it should be usable as a stand-alone binary in a windowless server environment as a component, and as a front-end based software that can leverage the binary of itself.
\section{Implementation of MASC}\label{sec:masc}
\begin{figure}[tbp]
	\centering
    \includegraphics[width=.47\textwidth]{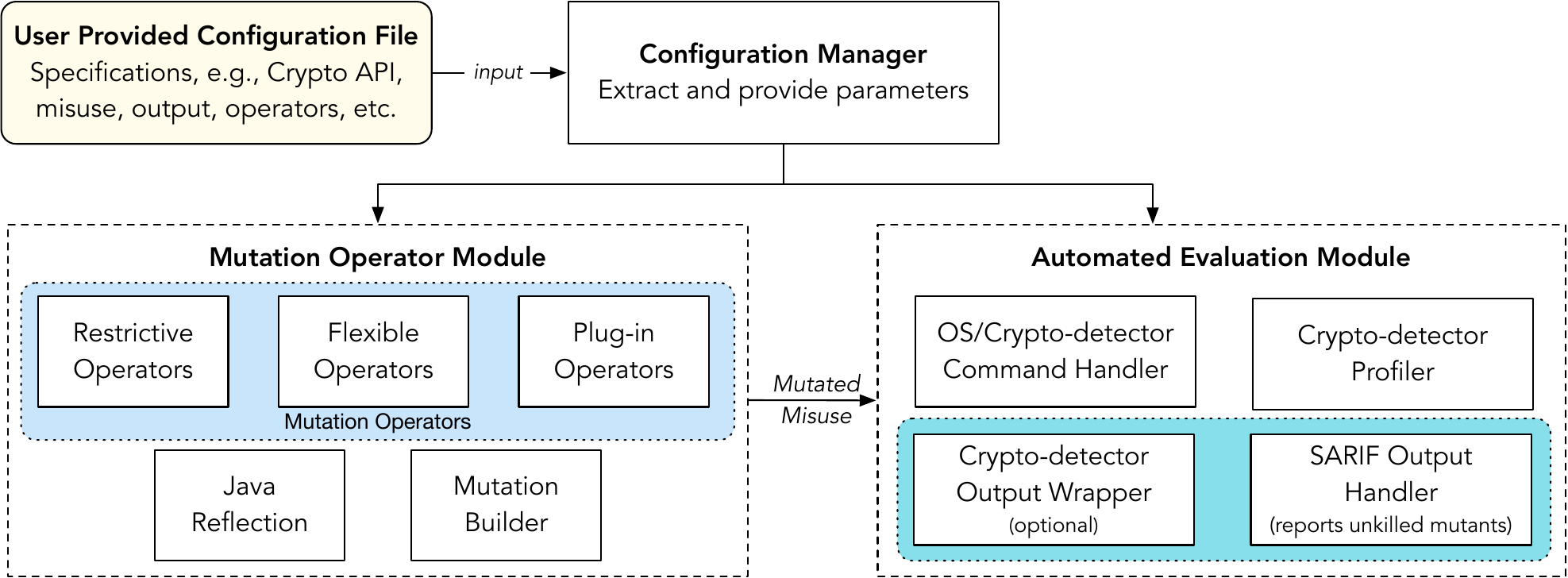}
    \caption{\small Architecture Overview of the Main Scope of \tool{}}
    \label{fig:tool-main-scope}
    \Description{Architecture overview of the Main scope of \tool{}. There are three high-level modules. Configuration Manager, the first module, receives the configuration values of \tool{} as a file from the user. Mutation Operator, the second module, creates mutated instances of crypto-API misuse and places them in the Main Scope. Finally, Automated Evaluation Module manages a \detector{} for automated evaluation. }
\end{figure}

To satisfy the design goals (\dgref{1}--\dgref{4}), we implemented \tool{} ($11K+$ effective Java source line of code) following single\-/responsibility principle across modules, classes, and functions.
Note that while current implementation of \tool{} inherits the \textit{mutation scopes} of the original implementation with internal structural changes, the bulk of the changes with new features in the current implementation of \tool{} are based on the \textit{Main Scope}.
Therefore, we describe the implementation details of \tool{} with a focus on \textit{Main Scope} in the context of the design goals and provide an overview of the architecture in Figure~\ref{fig:tool-main-scope}.

\myparagraphnew{Configuration Manager}
\begin{lstlisting}[basicstyle=\ttfamily\scriptsize,float,numbers=none,caption={{\small
    Example configuration file for \tool{}}},belowcaptionskip=-5mm,label=lst:example_config,emphstyle=\bfseries, language=bash]
scope = main
type = StringOperator
outputDir = app/outputs
apiName = javax.crypto.Cipher
# Method call from crypto-API
invocation = getInstance
# Secure parameter to use with crypto-API
secureParam = AES/GCM/NoPadding
# insecure parameter to use with crypto-API
insecureParam = AES
# noise value used with mutation
noise = ~
# variable, class name used to create necessary structures
variableName = cryptoVariable
className = CryptoTest
# name of the app for similarity-scope specific mutation
appName = <Name of the App>
\end{lstlisting}

To make \tool{} as flexible as possible, we decoupled the crypto-API specific parameters from the internal structure of \tool{}.
As a result, user can specify any crypto-API along with its necessary parameters through an external configuration file defining the base crypto-API misuse case. The configuration file follows a key-value format, as shown in Listing~\ref{lst:example_config}.
Additionally, user can specify the mutation operators and scope to be used, along with other configuration values, thus satisfying \dgref{1}.

\myparagraphnew{Mutation Operator Module}
\tool{} analyzes the specified crypto-API and uses the values specified by the user (\eg secure, and insecure parameters to be used with the API) for creating mutated crypto-API misuse instances.
Internally, the decoupling of crypto-APIs from \tool{} is made possible through the use of \textit{Java Reflection} based API analysis and Java Source Generation using the \textit{Java Poetry} Library (\dgref{1}).
While both the original and current implementation of \tool{} comes with several generalizable mutation operators, the current implementation of \tool{} includes an additional plug-in structure that facilitates creating custom mutation operators and custom key-value pairs for the configuration file.
Both of these can be done \textit{externally}, \ie no modification to source code of \tool{} is necessary (\dgref{2}). We provide additional details about \tools{} mutation operators in Section~\ref{sec:mutation-operators}.

\myparagraphnew{Automated Evaluation Module}
The current implementation of \tool{} leverages the SARIF formatted output to automate evaluation of \detectors{}.
To make end-to-end analysis automated, \tools{} can be configured to use \detector{} specific commands, such as \eg compiling a mutated source code for analysis, evaluation stop conditions, command for running \detector{}, output directory, and more (\dgref{3}--\dgref{4}).

Furthermore, \tool{} is implemented to produce verbose logs.
With the combination of flexible configuration, it is therefore possible to use the stand-alone binary \tool{} jar file as a module of another software. As a proof of concept, we implemented \mascweb{}, a \textit{python-django} based front-end that offers all the functionalities of the \tool{} (Usage details in Section~\ref{sec:using-masc}) that uses the binary jar of \tool{} as a module (\dgref{4}).

\subsection{Mutation Operators}\label{sec:mutation-operators}
We designed generalizable mutation operators
by examining the Java Cryptographic Architecture (JCA) documentation.
We identified two common patterns of crypto-API invocation as follows:
\bnum{i} \textit{restrictive}, where a developer is expected to only instantiate certain crypto-API objects by providing values from a pre-defined set, \eg{} \cipher, and \bnum{ii} \textit{flexible}, where the developers implement the behavior, \eg \hostnameVerifier{}.
While defining mutation operators of these two distinct patterns, we assumed a threat model consisting of the following types of adversaries:

\myparagraphnew{Benign developer, accidental misuse (T1)} A benign developer who accidentally misuses crypto-API, but attempts to address such vulnerabilities using a \detector{}.

\myparagraphnew{Benign developer, harmful fix (T2)} A benign developer who is trying to address a vulnerability identified by a \detector{} in good faith, but ends up introducing a new vulnerability instead.

\myparagraphnew{Evasive developer, harmful fix (T3)} A developer who aims to finish a task as quickly or with low effort (\eg a contractor), and is hence attempting to purposefully evade a \detector{}.

The restrictive operators mutate the restrictive values that abstracts away the crypto-API misuse.
For example, the abstraction can be based on method chaining, changing letter case (JCA is case-insensitive), and introducing alias variables, as shown in Listing~\ref{lst:mutations}.
We implemented $6$ mutation operators for restrictive crypto-APIs.
\newline
\noindent Similarly, for the flexible APIs, we implemented mutation operators based on object-oriented programming concepts:
\begin{itemize}[leftmargin=*]
    \item\textbf{Method overriding} is used to create mutations that contain \textit{ineffective} security exception statements, irrelevant loops, and/or ineffective security sensitive return value,
    \item\textbf{Class extension} is used for implementing or inheriting parent crypto-API interface or abstract classes respectively, and
    \item\textbf{Object Instantiation} is for creating anonymous inner class object from the implemented or inherited classes of crypto-APIs.
\end{itemize}
\begin{lstlisting}[basicstyle=\ttfamily\scriptsize,float,numbers=none,caption={{\small
    Flexible crypto-API based misuse mutation by \tool{}}},belowcaptionskip=-5mm,label=lst:flexible,emph={mutation},emphstyle=\bfseries, language=java]
interface IHV extends HostnameVerifier{}
new IHV(){
 public boolean verify(String h,SSLSession s)return true;};
\end{lstlisting}

We created $6$ more conceptual mutation operators based on flexible crypto-APIs. An example of flexible mutant is shown in Listing~\ref{lst:flexible}.

\subsection{Mutation Scopes}\label{sec:mutation-scopes}
To emulate vulnerable crypto-API misuse placement by benign and evasive developers, we designed three mutation scopes to be used with \tool{}:
\begin{itemize}[leftmargin=*]
	\item \textit{Main Scope} represents the simplest scope, where it seeds mutants at the beginning of the main method of a simple Java or Android template app, ensuring reachability.

	\item \textit{Similarity Scope}, which is extended from \mdroid~\cite{mtb+18,lbt+17}, seeds mutants in the source code of an input application where a similar crypto-API is found. Note that it does not modify the existing crypto-API, and only appends the said mutant misuse case

	\item \textit{Exhaustive Scope}, which is extended \mse{}~\cite{AKM+21,demo-muse-2021,bkm+18}, seeds mutants at \textit{all syntactically possible} locations in the target app, such as class definition, conditional segments, method bodies and anonymous inner class object declarations. This helps evaluate the reachability of the target \detector{}.
\end{itemize}

\section{Using MASC}\label{sec:using-masc}
\begin{figure}[tbp]
    \centerline{\includegraphics[width=0.48\textwidth]{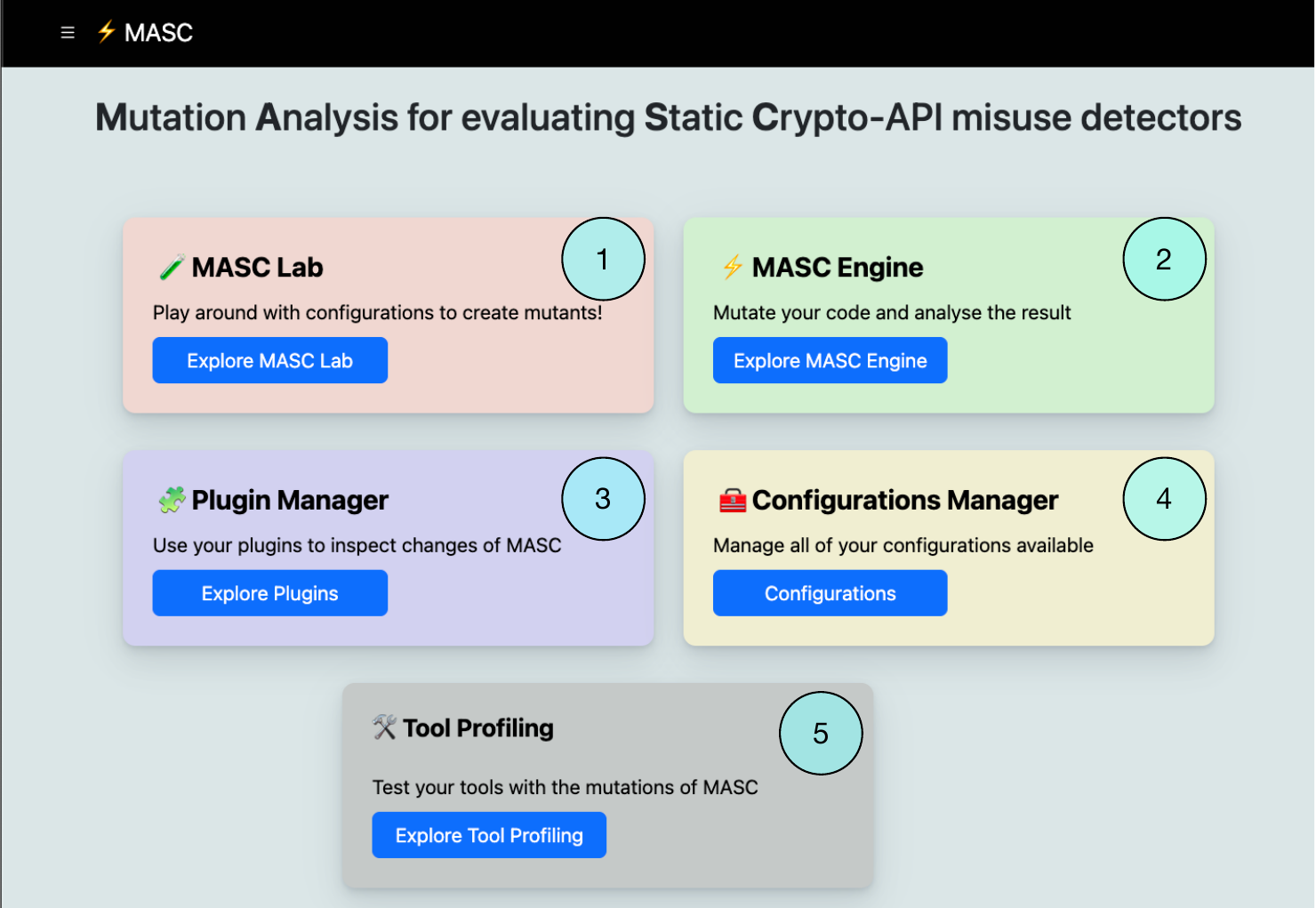}}
    \caption{\small Web based Front-end of the \tool{}}\label{fig:masc-front-end}
\end{figure}
As described previously, \tool{} has both command line interface and web-based front-end (\mascweb{}, shown in Figure~\ref{fig:masc-front-end}). \tool{} CLI can be executed by providing a configuration file \eg Cipher.properties using the command shown in Listing~\ref{lst:climasc}.
\begin{lstlisting}[basicstyle=\ttfamily\scriptsize,float,numbers=none,caption={{\small
    Running \tool{} CLI with a configuration file}},belowcaptionskip=-5mm,label=lst:climasc,emphstyle=\bfseries, language=bash]
java -jar MASC.jar Cipher.properties
\end{lstlisting}

Similarly, using the \mascweb, users can do the following, labeled as per Figure~\ref{fig:masc-front-end}:
\begin{enumerate}[leftmargin=*]
\item Experiment and learn about crypto-API misuse using \masclab{},
\item Mutate open source applications by uploading the zipped source code in \mascengine{},
\item Use custom implemented mutation operators as plugins,
\item Create and upload configuration files, and
\item Profile \detectors{} by analyzing caught and uncaught mutants.
\end{enumerate}
The detailed description of each of these, with example configuration files, and detailed developer documentation, is shared in the open-source repository of \tool~\cite{MASC}.
\vspace{.4em}
\section{Future Work and Conclusion}
\label{sec:conclusion}
We discussed the overview, design goals, implementation details and usage of \tool{}, a user-friendly tool for mutation-based evaluation of static crypto-API misuse detectors.
While we do not report any additional \detector evaluation in this demonstration paper, evaluation results of the original implementation of \tool{} are available in the original paper~\cite{ami-masc-oakland22}.
We plan to evaluate additional \detectors with the current implementation of \tool{}, and aim to extend the customization support to the additional scopes, \ie \exhaustiveScope{} and \similarityScope{}.
We hope that the current implementation of \tool{} will help \detector stakeholders, \ie security researchers, developers and users, to systematically evaluate \detectors{}.
Furthermore, we envision that that open-source enthusiasts will augment the mutation operators of \tool{} further, empowered by its easy to extend architecture, thus helping improve \detectors{} by finding novel flaws.

\begin{acks}
  This work is supported in part by NSF-$1815336$, NSF-$1815186$, NSF-$1955853$ grants and Coastal Virginia Center for Cyber Innovation and the Commonwealth Cyber Initiative, an investment in the advancement of cyber R\&D, innovation, and workforce development. For more information about COVA CCI and CCI, visit www.covacci.org and www.cyberinitiative.org.
\end{acks}

\bibliographystyle{ACM-Reference-Format}
\bibliography{extracted.bib}


\begin{thebibliography}{18}


\ifx \showCODEN    \undefined \def \showCODEN     #1{\unskip}     \fi
\ifx \showDOI      \undefined \def \showDOI       #1{#1}\fi
\ifx \showISBNx    \undefined \def \showISBNx     #1{\unskip}     \fi
\ifx \showISBNxiii \undefined \def \showISBNxiii  #1{\unskip}     \fi
\ifx \showISSN     \undefined \def \showISSN      #1{\unskip}     \fi
\ifx \showLCCN     \undefined \def \showLCCN      #1{\unskip}     \fi
\ifx \shownote     \undefined \def \shownote      #1{#1}          \fi
\ifx \showarticletitle \undefined \def \showarticletitle #1{#1}   \fi
\ifx \showURL      \undefined \def \showURL       {\relax}        \fi
\providecommand\bibfield[2]{#2}
\providecommand\bibinfo[2]{#2}
\providecommand\natexlab[1]{#1}
\providecommand\showeprint[2][]{arXiv:#2}

\bibitem[MAS(2022)]%
        {MASCARTIFACT}
Secure Platforms Lab \bibinfo{year}{2022}\natexlab{}.
\newblock \bibinfo{booktitle}{\emph{{MASC Artifact}}}.
\newblock Secure Platforms Lab.
\newblock
\urldef\tempurl%
\url{https://github.com/Secure-Platforms-Lab-W-M/MASC-Artifact}
\showURL{%
Retrieved May, 2023 from \tempurl}


\bibitem[MAS(2023)]%
        {MASC}
Secure Platforms Lab \bibinfo{year}{2023}\natexlab{}.
\newblock \bibinfo{booktitle}{\emph{{MASC}}}.
\newblock Secure Platforms Lab.
\newblock
\urldef\tempurl%
\url{https://github.com/Secure-Platforms-Lab-W-M/MASC}
\showURL{%
Retrieved May, 2023 from \tempurl}


\bibitem[Ami et~al\mbox{.}(2022)]%
        {ami-masc-oakland22}
\bibfield{author}{\bibinfo{person}{{Amit Seal} Ami}, \bibinfo{person}{Nathan
  Cooper}, \bibinfo{person}{Kaushal Kafle}, \bibinfo{person}{Kevin Moran},
  \bibinfo{person}{Denys Poshyvanyk}, {and} \bibinfo{person}{Adwait Nadkarni}.}
  \bibinfo{year}{2022}\natexlab{}.
\newblock \showarticletitle{{Why Crypto-detectors Fail: A Systematic Evaluation
  of Cryptographic Misuse Detection Techniques}}. In
  \bibinfo{booktitle}{\emph{2022 IEEE Symposium on Security and Privacy
  (S\&P)}}. \bibinfo{publisher}{IEEE Computer Society}, \bibinfo{address}{San
  Francisco, CA, USA}, \bibinfo{pages}{397--414}.
\newblock
\showISSN{2375-1207}
\urldef\tempurl%
\url{https://doi.org/10.1109/SP46214.2022.9833582}
\showDOI{\tempurl}


\bibitem[Ami et~al\mbox{.}(2021a)]%
        {demo-muse-2021}
\bibfield{author}{\bibinfo{person}{{Amit Seal} Ami}, \bibinfo{person}{Kaushal
  Kafle}, \bibinfo{person}{Kevin Moran}, \bibinfo{person}{Adwait Nadkarni},
  {and} \bibinfo{person}{Denys Poshyvanyk}.} \bibinfo{year}{2021}\natexlab{a}.
\newblock \showarticletitle{{Demo: Mutation-based Evaluation of
  Security-focused Static Analysis Tools for Android}}. In
  \bibinfo{booktitle}{\emph{{Proceedings of the 43rd IEEE/ACM International
  Conference on Software Engineering (ICSE’21), Formal Tool Demonstration,
  Virtual (originally Madrid, Spain), May 25th - 28th, 2021}}}.
\newblock


\bibitem[Ami et~al\mbox{.}(2021b)]%
        {AKM+21}
\bibfield{author}{\bibinfo{person}{Amit~Seal Ami}, \bibinfo{person}{Kaushal
  Kafle}, \bibinfo{person}{Kevin Moran}, \bibinfo{person}{Adwait Nadkarni},
  {and} \bibinfo{person}{Denys Poshyvanyk}.} \bibinfo{year}{2021}\natexlab{b}.
\newblock \showarticletitle{Systematic {{Mutation}}-{{Based Evaluation}} of the
  {{Soundness}} of {{Security}}-{{Focused Android Static Analysis
  Techniques}}}.
\newblock \bibinfo{journal}{\emph{ACM Transactions on Privacy and Security}}
  \bibinfo{volume}{24}, \bibinfo{number}{3} (\bibinfo{date}{Feb.}
  \bibinfo{year}{2021}), \bibinfo{pages}{15:1--15:37}.
\newblock
\showISSN{2471-2566}
\urldef\tempurl%
\url{https://doi.org/10.1145/3439802}
\showDOI{\tempurl}


\bibitem[Bessey et~al\mbox{.}(2010)]%
        {BBC+10}
\bibfield{author}{\bibinfo{person}{Al Bessey}, \bibinfo{person}{Ken Block},
  \bibinfo{person}{Ben Chelf}, \bibinfo{person}{Andy Chou},
  \bibinfo{person}{Bryan Fulton}, \bibinfo{person}{Seth Hallem},
  \bibinfo{person}{Charles {Henri-Gros}}, \bibinfo{person}{Asya Kamsky},
  \bibinfo{person}{Scott McPeak}, {and} \bibinfo{person}{Dawson Engler}.}
  \bibinfo{year}{2010}\natexlab{}.
\newblock \showarticletitle{A {{Few Billion Lines}} of {{Code Later}}: {{Using
  Static Analysis}} to {{Find Bugs}} in the {{Real World}}}.
\newblock \bibinfo{journal}{\emph{Commun. ACM}} \bibinfo{volume}{53},
  \bibinfo{number}{2} (\bibinfo{date}{Feb.} \bibinfo{year}{2010}),
  \bibinfo{pages}{66--75}.
\newblock
\showISSN{0001-0782}
\urldef\tempurl%
\url{https://doi.org/10.1145/1646353.1646374}
\showDOI{\tempurl}


\bibitem[Bonett et~al\mbox{.}(2018)]%
        {bkm+18}
\bibfield{author}{\bibinfo{person}{Richard Bonett}, \bibinfo{person}{Kaushal
  Kafle}, \bibinfo{person}{Kevin Moran}, \bibinfo{person}{Adwait Nadkarni},
  {and} \bibinfo{person}{Denys Poshyvanyk}.} \bibinfo{year}{2018}\natexlab{}.
\newblock \showarticletitle{Discovering Flaws in Security-Focused Static
  Analysis Tools for {Android} using Systematic Mutation}. In
  \bibinfo{booktitle}{\emph{27th {USENIX} Security Symposium ({USENIX} Security
  18)}}. \bibinfo{publisher}{{USENIX} Association},
  \bibinfo{address}{Baltimore, MD}, \bibinfo{pages}{1263--1280}.
\newblock
\showISBNx{978-1-939133-04-5}
\urldef\tempurl%
\url{https://www.usenix.org/conference/usenixsecurity18/presentation/bonett}
\showURL{%
\tempurl}


\bibitem[CogniCrypt(2020)]%
        {cognicrypteclipse}
\bibfield{author}{\bibinfo{person}{CogniCrypt}.}
  \bibinfo{year}{2020}\natexlab{}.
\newblock \bibinfo{booktitle}{\emph{{{CogniCrypt}} - {{Secure Integration}} of
  {{Cryptographic Software}} | {{CogniCrypt}}}}.
\newblock
\urldef\tempurl%
\url{{https://www.eclipse.org/cognicrypt/}}
\showURL{%
\tempurl}
\newblock
\shownote{Accessed June, 2020}.


\bibitem[CryptoGuard(2020)]%
        {cryptoguard_oracle}
\bibfield{author}{\bibinfo{person}{CryptoGuard}.}
  \bibinfo{year}{2020}\natexlab{}.
\newblock \bibinfo{booktitle}{\emph{{Oracle - Industrial Experience of Finding
  Cryptographic Vulnerabilities in Large-scale Codebases}}}.
\newblock
\urldef\tempurl%
\url{{https://labs.oracle.com/pls/apex/f?p=94065:40150:0::::P40150_PUBLICATION_ID:6629}}
\showURL{%
\tempurl}
\newblock
\shownote{Accessed July, 2020}.


\bibitem[GitHub(2020)]%
        {github_third_party_code_scanners}
\bibfield{author}{\bibinfo{person}{GitHub}.} \bibinfo{year}{2020}\natexlab{}.
\newblock \bibinfo{booktitle}{\emph{{{Announcing third-party code scanning
  tools: static analysis \& developer security training - The GitHub Blog}}}}.
\newblock
\urldef\tempurl%
\url{{https://github.blog/2020-10-05-announcing-third-party-code-scanning-tools-static-analysis-and-developer-security-training/}}
\showURL{%
\tempurl}
\newblock
\shownote{Accessed Nov, 2020}.


\bibitem["Java"(2020)]%
        {JavaCryptographyArchitecture}
\bibfield{author}{\bibinfo{person}{"Java"}.} \bibinfo{year}{2020}\natexlab{}.
\newblock \bibinfo{title}{Java {{Cryptography Architecture}} ({{JCA}})
  {{Reference Guide}}}.
\newblock
\newblock
\urldef\tempurl%
\url{https://docs.oracle.com/en/java/javase/11/security/java-cryptography-architecture-jca-reference-guide.html\#GUID-815542FE-CF3D-407A-9673-CAE9840F6231}
\showURL{%
\tempurl}


\bibitem[lgtm(2020)]%
        {lgtm}
\bibfield{author}{\bibinfo{person}{lgtm}.} \bibinfo{year}{2020}\natexlab{}.
\newblock \bibinfo{title}{{{LGTM}} - {{Continuous}} Security Analysis}.
\newblock
\newblock
\urldef\tempurl%
\url{{https://lgtm.com/}}
\showURL{%
\tempurl}
\newblock
\shownote{Accessed Nov, 2020}.


\bibitem[{Linares-V{\'a}squez} et~al\mbox{.}(2017)]%
        {lbt+17}
\bibfield{author}{\bibinfo{person}{Mario {Linares-V{\'a}squez}},
  \bibinfo{person}{Gabriele Bavota}, \bibinfo{person}{Michele Tufano},
  \bibinfo{person}{Kevin Moran}, \bibinfo{person}{Massimiliano Di~Penta},
  \bibinfo{person}{Christopher Vendome}, \bibinfo{person}{Carlos
  {Bernal-C{\'a}rdenas}}, {and} \bibinfo{person}{Denys Poshyvanyk}.}
  \bibinfo{year}{2017}\natexlab{}.
\newblock \showarticletitle{Enabling {{Mutation Testing}} for {{Android
  Apps}}}. In \bibinfo{booktitle}{\emph{Proceedings of the 2017 11th {{Joint
  Meeting}} on {{Foundations}} of {{Software Engineering}}}}
  \emph{(\bibinfo{series}{{{ESEC}}/{{FSE}} 2017})}.
  \bibinfo{publisher}{{Association for Computing Machinery}},
  \bibinfo{address}{{New York, NY, USA}}, \bibinfo{pages}{233--244}.
\newblock
\showISBNx{978-1-4503-5105-8}
\urldef\tempurl%
\url{https://doi.org/10.1145/3106237.3106275}
\showDOI{\tempurl}


\bibitem[Moran et~al\mbox{.}(2018)]%
        {mtb+18}
\bibfield{author}{\bibinfo{person}{Kevin Moran}, \bibinfo{person}{Michele
  Tufano}, \bibinfo{person}{Carlos {Bernal-C{\'a}rdenas}},
  \bibinfo{person}{Mario {Linares-V{\'a}squez}}, \bibinfo{person}{Gabriele
  Bavota}, \bibinfo{person}{Christopher Vendome}, \bibinfo{person}{Massimiliano
  Di~Penta}, {and} \bibinfo{person}{Denys Poshyvanyk}.}
  \bibinfo{year}{2018}\natexlab{}.
\newblock \showarticletitle{{{MDroid}}+: {{A Mutation Testing Framework}} for
  {{Android}}}.
\newblock \bibinfo{journal}{\emph{Proceedings of the 40th International
  Conference on Software Engineering Companion Proceeedings - ICSE '18}}
  (\bibinfo{year}{2018}), \bibinfo{pages}{33--36}.
\newblock
\urldef\tempurl%
\url{https://doi.org/10.1145/3183440.3183492}
\showDOI{\tempurl}


\bibitem[OASIS(2021)]%
        {SARIF}
\bibfield{author}{\bibinfo{person}{OASIS}.} \bibinfo{year}{2021}\natexlab{}.
\newblock \bibinfo{title}{{{The Static Analysis Results Interchange Format
  (SARIF)}}}.
\newblock
\newblock
\urldef\tempurl%
\url{{https://sarifweb.azurewebsites.net/ }}
\showURL{%
\tempurl}
\newblock
\shownote{Accessed Jul, 2021}.


\bibitem[owasp(2020)]%
        {owasp:mislabel}
\bibfield{author}{\bibinfo{person}{owasp}.} \bibinfo{year}{2020}\natexlab{}.
\newblock \bibinfo{title}{Test Cases for Risky or Broken Cryptographic
  Algorithm Erroneously Labeled as Not Vulnerable {$\cdot$} {{Issue}} \#92
  {$\cdot$} {{OWASP}}/{{Benchmark}}}.
\newblock
\newblock
\urldef\tempurl%
\url{{https://github.com/OWASP/Benchmark/issues/92}}
\showURL{%
\tempurl}
\newblock
\shownote{Accessed Nov, 2020}.


\bibitem[Rahaman et~al\mbox{.}(2019)]%
        {rxa+19}
\bibfield{author}{\bibinfo{person}{Sazzadur Rahaman}, \bibinfo{person}{Ya
  Xiao}, \bibinfo{person}{Sharmin Afrose}, \bibinfo{person}{Fahad Shaon},
  \bibinfo{person}{Ke Tian}, \bibinfo{person}{Miles Frantz},
  \bibinfo{person}{Murat Kantarcioglu}, {and} \bibinfo{person}{Danfeng~(Daphne)
  Yao}.} \bibinfo{year}{2019}\natexlab{}.
\newblock \showarticletitle{{CryptoGuard: High Precision Detection of
  Cryptographic Vulnerabilities in Massive-sized Java Projects}}. In
  \bibinfo{booktitle}{\emph{Proceedings of the 2019 {{ACM SIGSAC Conference}}
  on {{Computer}} and {{Communications Security}} - {{CCS}} '19}}.
  \bibinfo{publisher}{{ACM Press}}, \bibinfo{address}{{London, United
  Kingdom}}, \bibinfo{pages}{2455--2472}.
\newblock
\showISBNx{978-1-4503-6747-9}
\urldef\tempurl%
\url{https://doi.org/10.1145/3319535.3345659}
\showDOI{\tempurl}


\bibitem[Xanitizer(2020)]%
        {XanitizerRIGSIT}
\bibfield{author}{\bibinfo{person}{Xanitizer}.}
  \bibinfo{year}{2020}\natexlab{}.
\newblock \bibinfo{title}{Xanitizer by {{RIGS IT}} - {{Because Security
  Matters}}}.
\newblock
\newblock
\urldef\tempurl%
\url{{{https://www.rigs-it.com/xanitizer/}}}
\showURL{%
\tempurl}
\newblock
\shownote{Accessed May, 2020}.


\end{thebibliography}

\end{document}